\documentclass[twocolumn]{revtex4-1}
\usepackage{graphicx}

\begin{document}

\title{Spin models as microfoundation of macroscopic financial market models} 
\author{Sebastian M.\ Krause}
\author{Stefan Bornholdt}
\affiliation{Institut f\"ur Theoretische Physik, Universit\"at Bremen, Otto-Hahn-Allee 1, 28359 Bremen}
\begin{abstract}
Macroscopic price evolution models are commonly used for investment strategies. There are first promising achievements in defining microscopic agent based models for the same purpose. Microscopic models allow a deeper understanding of mechanisms in the market than the purely phenomenological macroscopic models, and thus bear the chance for better models for market regulation. We exemplify this strategy in a case study, deducing a macroscopic Langevin equation from a microscopic spin market model closely related to the Ising model. The interplay of the microscopic and the macroscopic view allows for a better understanding of the microscopic model, as well, and may guide the construction of agent based market models as basis of macroscopic price models. 
\end{abstract}
\maketitle

While extensively used textbook models of financial markets assume an efficient market \cite{fama1970} and Gaussian distributed returns of financial assets \cite{bachelier1900,black1973}, real markets show a completely different behavior with huge bubbles, crashes, and erratic asset price time series. This discrepancy is believed to be one of the central reasons for the 2008 financial market crisis \cite{lux2009}. Scientific investigations beyond the standard theories of financial markets are partly driven by methods from theoretical physics, especially in the description and modeling of macroeconomic quantities (e.g.\ price time series) as well as in the creation of microeconomic models with many interacting agents \cite{farmer2009}. 

Price time series of markets often exhibit universal properties, known as stylized facts, that do not occur in equilibrium models of financial markets: Returns are power law distributed \cite{mandelbrot1963} and their signs are uncorrelated, volatility occurs clustered \cite{gopikrishnan1999}, and multi-fractal properties occur \cite{ding1993}. The first three stylized facts have been incorporated in different volatility forecasting models (broadly used for investment strategies) like ARMA and GARCH variants (for an overview and comparison see \cite{poon2003}). A recent development also includes multi-fractal properties with reasonable success \cite{calvet2001,lux2008}. 
Microeconomic studies have lead to a number of agent models (see \cite{samanidou2007} for an overview) which, however, are mostly in the stage of anecdotal models. As an early model, the Lux Marchesi model \cite{lux1999} introduces chartist agents competing with fundamentalists, leading to power law distributed returns as observed in real markets, contradicting the popular efficient market hypothesis. A minimalistic spin model \cite{bornholdt2001} introduces spatial structure and exhibits pronounced bull and bear markets with a long memory as in real markets \cite{kaizoji2002}.

As proposed by Farmer \cite{farmer2009} and exhibited in \cite{gatti2007,wiesinger2008}, agent based models can be brought so close to real markets that they are candidates for investment strategy tools. While in \cite{gatti2007,wiesinger2008} the outcome of agent based models was directly compared to price time series, we here deduce a macroscopic price equation from the microscopic market model of ref.\ \cite{bornholdt2001} which is based on a modified Ising spin model. Macro-relations for agent based models are discussed in \cite{ramsey1996,lux1997,pakkanen2010} and suggested for the spin model \cite{bornholdt2001} in \cite{giardina2003}. There are several advantages of this procedure: The focus shift from macro to micro-models is facilitated, the interplay of microscopic and macroscopic views assists agent based model construction, and finally with multi-fractal properties the micro- and the macro-representation might share the same principles. 

We here focus on the spin model defined by Bornholdt \cite{bornholdt2001}, since it allows for a clear geometric insight, connects to the very well known Ising model, and has been studied as a minimalistic agent based model exhibiting stylized facts \cite{kaizoji2002,yamamoto2010}. We first analyze the dynamics of the model with a frozen coupling to the global magnetization of the system and find a phase transition between small and large magnetizations with a jump in volatility. Such a phase transition is found in other socio-economic models, as well \cite{harras2010}. Then we construct a macroscopic equation for the price, which includes a volatility switching due to the phase transition, reminiscent of regime switching models \cite{poon2003}. The macroscopic model is compared to the microscopic one with very good agreement and thus provides insight in the mechanism ruling the spin market model. From the macroscopic point of view we investigate the role of the model parameters. Finally, we propose the use of deduced macro-equations as a guiding principle for the construction of agent based models. 

The spin model of ref.\ \cite{bornholdt2001} describes $N=L^2$ agents on a square lattice with periodic boundary conditions. The agents can decide between two states $S_i=\pm 1$ (buy or sell), according to two conflicting forces: On the one hand they tend to do what their neighbors do (as in the Ising model), while on the other hand they tend to join the minority, if minority and majority sizes diverge (as in the minority game \cite{minoritygame}). This is implemented by a local field 
\begin{eqnarray}
h_i & = & \sum_{j \ \in \rm nn(i)}S_j - |h|\cdot s_i \label{eq:locfield}\\
h & = & \alpha \cdot m = \alpha \cdot \frac{1}{N}\sum_{k=1}^N S_k \label{eq:meanfield}
\end{eqnarray}
with nn(i) denoting the nearest neighbors of agent $i$. In random sequential update, a randomly chosen agent $i$ evolves according to the local field $h_i$. Its new state $S_i$ is determined by the probabilities 
\begin{eqnarray*}
S_i = +1 & {\quad {\rm with}\quad} & p_{\rm u}=\frac{1}{1+\exp(-2\beta h_i)}  \\
S_i = -1 & {\quad {\rm with}\quad} & p_{\rm d}=1-p_{\rm u}.
\end{eqnarray*}
For $\alpha=0$ this corresponds to a simulation of the pure Ising model with a heat bath algorithm and coupling $J=1$. If $\alpha \neq 0$, the system couples to its mean field $h$ as a time-dependent global quantity (comparable to the well known global property ``price" in economy). Note that this coupling is of an entirely different type than the coupling of a spin to an external magnetic field, the latter being described by a term $-h$ instead of $-|h|\cdot S_i$. This results in a much more complex behavior compared to the Ising model. The time dependence of $h$ introduces a feedback to the system's dynamics. Later on we will fix $h$ to certain constant values in order to understand the role of $h(t)$. 

\begin{figure}[tb]
\includegraphics[width=1.0\columnwidth]{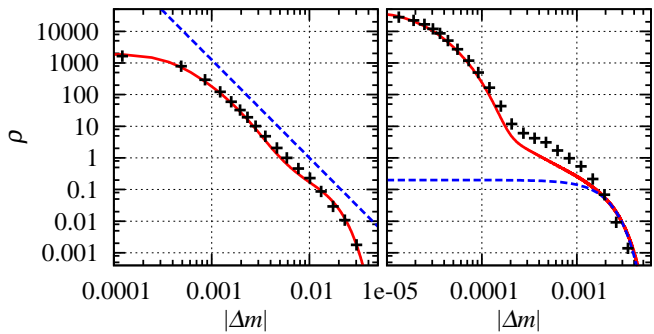}
\caption{\label{fig:returns}Comparison of the absolute return densities $\rho(|\Delta m|)$ provided by the microscopic (crosses) and the macroscopic model (lines). Left: $L=128,\ \alpha=20$ and a power law with exponent $-3.1$ (dashed line). Right: $L=1024,\ \alpha=60$ and a Gaussian with $\sigma=0.0012$ (dashed line).}
\end{figure}

The dynamics of this model exhibits large and broadly distributed fluctuations in the magnetization $m$ that we here relate to fluctuations in financial markets. For this purpose, price has been introduced via additional fundamentalists, leading to an expression for logarithmic returns, which is identified with the change of magnetization in the model, ${\rm ret}(t)=\ln(p(t)/p(t-1))\propto \Delta m$ \cite{kaizoji2002}. As can be seen in Fig.~\ref{fig:returns} on the left, where the absolute return distribution of the model is shown with symbols, the returns are distributed with a truncated power law and an exponent of approximately 3 (or exponent 2 for the cumulative distribution). In addition, a long-range autocorrelation as well as multi-fractal properties are reported \cite{kaizoji2002}. 

\begin{figure}[tb]
\includegraphics[width=1.0\columnwidth]{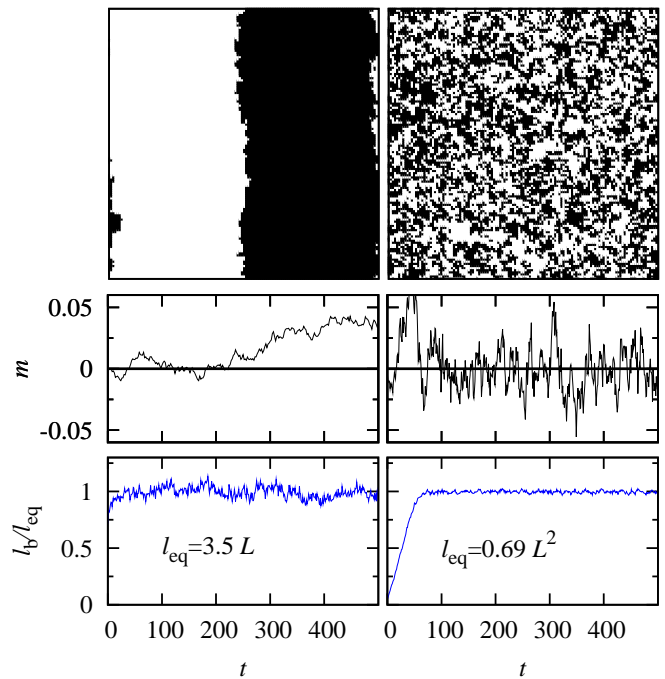}
\caption{\label{fig:frozen}From top to bottom: Lattice states ($L=128$), magnetization $m$, and borderline length time series for $h=1.5$ (left column) and $h=2.0$ (right column).}
\end{figure}

\begin{figure}[tb]
\includegraphics[width=1.0\columnwidth]{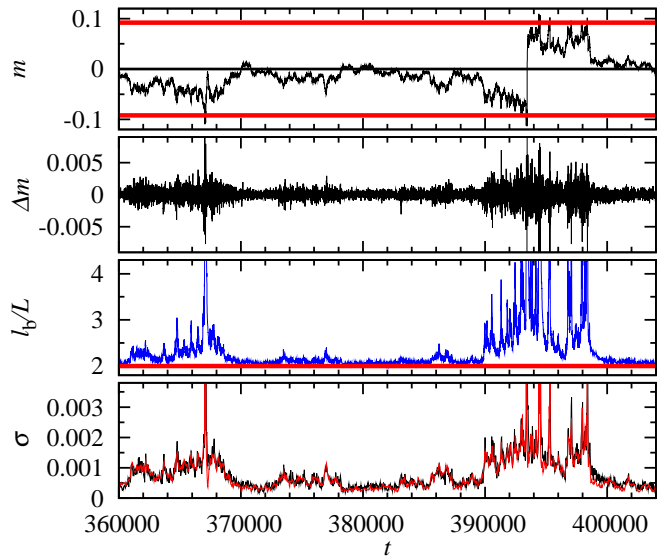}
\caption{\label{fig:volat}From top to bottom: Time series of magnetization $m$ within the borders of $\pm m_{\rm crit}$; returns $\Delta m$; border length and minimal border length allowing for $m \approx 0$; volatility calculated as $\bar{\sigma}(t)$ (black) and as $\sigma(L,\alpha\bar{m}(t))$ (red). The parameters are set to $L=128$ and $\alpha=20$.}
\end{figure}

In order to characterize the lattice state, the border line length 
\begin{eqnarray}
l_{\rm b} & = & \sum_{\left<ij\right>} \Theta(-S_i S_j) \label{eq:lengthb}
\end{eqnarray}
is introduced, where $\left<ij\right>$ are pairs of nearest neighbors and $\Theta(1)=1,\ \Theta(-1)=0$. This property is related to the number of agents, which are exposed to stress due to different neighbors and hence switch more often (The other agents switch with a probability close to zero. On a side note, this allows for a faster algorithm scaling with $L$ instead of $L^2$.). Most of the time it is near $2L$ (compare Fig.~\ref{fig:volat}) indicating that the system is undercritical most of the time (compare Fig.~\ref{fig:frozen}, upper left).

To simplify the dynamics, let us now set the mean field $h$ in eq.~(\ref{eq:locfield}) to a constant value (i.e.\ treat it as an external constant field). Because the dynamics of the spin model leads to states with rectangular components most of the time, we use $S_i=1$ for $x<\frac{L}{2}$ and $S_i=-1$ for $x\geq \frac{L}{2}$ as the initial state.
Figure \ref{fig:frozen} shows the state of the lattice (for the fixed time $t=400$ after equilibration) and some time-dependent quantities for a small (left column) and a large $h$ (right column). In this paper, the temperature is set to $0.2 \  T_c$ with  $T_c$ being the critical temperature of the Ising model. 

In the case with a small constant value of $h$ (typical in this model), the returns $\Delta m$ are Gaussian distributed and not correlated (also higher moments of $\Delta m$ are not correlated). Therefore we conclude that the magnetization performs a random walk. This can be modeled by 
\begin{eqnarray}
\Delta m (t) & = & \sigma (L,h) \cdot \xi_t \label{eq:langevine}
\end{eqnarray}
with the normally distributed and independent random variable $\xi_t$. Eq.~(\ref{eq:langevine}) is a macroscopic equation, since it deals with the global quantity $m$, only. 

For larger $h$, we get a completely different behavior, indicating a phase transition. The system is disordered globally and $l_{\rm b}$ is much larger (scaling with $L^2$ instead of $L$). $\Delta m$ is much larger, as well, and it is strongly anti-correlated since $m$ fluctuates fast. 
\begin{figure}[tb]
\includegraphics[width=1.0\columnwidth]{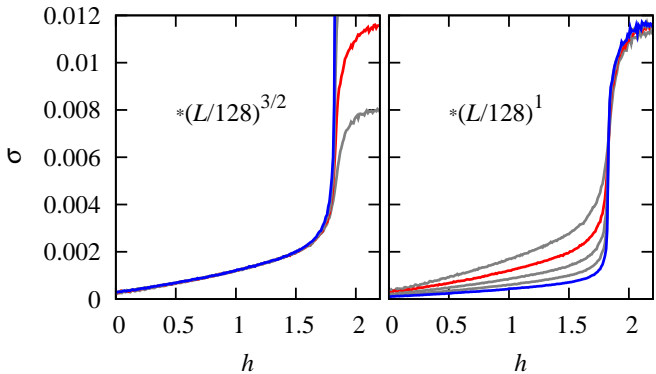}
\caption{\label{fig:transition}Averaged volatility $\sigma$ for fixed values of the field $h$ below and above the phase transition. $L=128$ (red), $L=1024$ (blue) and $L=64,\ 256,\ 512$ (grey).}
\end{figure}
Fig.~\ref{fig:transition} shows the volatility $\sigma$ as a function of $h$ for different $L$, calculated as the average over 10000 time steps after equilibration. Additional boundary conditions were used in order to avoid states with $m=\pm 1$: All agents at $x=L/4$ are fixed to $S_i=1$ (at $x=3L/4$ to $S_i=-1$). On the left, the data is multiplied with $\left(L/128\right)^{3/2}$. For the ordered phase (small $h$), then the curves collapse for different $L$ indicating a finite size scaling with $L^{-3/2}$. On the right hand side, the disordered part of the curves collapses with a scaling $\propto L^{-1}$. The phase transition is close to $h_{\rm crit}=1.86$ and becomes more pronounced with increasing $L$.

The reason for the scaling properties can be understood as follows: The number of active agents (those with much higher switching probabilities) is proportional to $l_{\rm b}$. So $l_{\rm b}$ can be seen as the number of steps contributing to a single return, and thus the total change expressed in steps scales with $l_{\rm b}^{1/2}$, if the single steps are treated as independent. Including the pre-factor of $1/L^2$ in the magnetization and with the scaling properties of $l_{\rm b}$ (compare Fig.~\ref{fig:frozen}) the observed scaling is found. 

Since in the spin market model the system is in the ordered phase most of the time, a macroscopic description with a generalization of eq.~(\ref{eq:langevine}) is promising. The only change to do is to let $h$ be time dependent. The time dependence should be smooth, so let us use a smoothened field $\bar{h}(t)=\alpha \bar{m}(t)$ and obtain
\begin{eqnarray}
\Delta m (t) & = & \sigma \left(L, \alpha \bar{m}(t) \right) \cdot \xi_t \label{eq:macro}
\end{eqnarray}
as an equation describing the price evolution as a macroscopic quantity. The dependence $\sigma(L,h)$ is taken from fits to the curves in Fig.~\ref{fig:transition}. Eq.~(\ref{eq:macro}) might be compared to macroscopic models derived from market data \cite{poon2003}, which propose mechanisms for the generation of $\sigma(t)$ time series with volatility clustering to reproduce real market's behavior. Here, via the frozen mean field, the macroscopic equation is deduced from a microscopic model as $\sigma(L,\bar{h}=\alpha\bar{m})$. It is worth noting that macroscopic models are usually constructed without using microscopic first principles, just with the goal of a good tunability and forecasting power. Multi-fractal models, on the other hand,  include insights into the market process itself \cite{calvet2001,lux2008} and thus can serve as examples for agent based model design, as well. 

Eq.~(\ref{eq:macro}) is valid only for fields $\bar{h}=\alpha \bar{m}$ beyond the phase transition, because in the disordered phase anti-correlation takes place and the equilibration process is not taken into account. We can put it as follows: The system's magnetization performs a random walk described by eq.~(\ref{eq:macro}) and additionally experiences the effect of complex and finally reflecting borders around $m_{\rm crit}=h_{\rm crit}/\alpha$ as can be seen in Fig.~\ref{fig:volat} on top. Since here we will not describe the action of the borders at $m_{\rm crit}$, we will use eq.~(\ref{eq:macro}) for supercritical $m$ too and use the time series $h(t)$ produced by the microscopic model for the further investigation of the macroscopic model. The jump in volatility due to the phase transition has an analogue in regime switching models \cite{poon2003}. 
The bottom panel of Fig.~\ref{fig:volat} shows the comparison of the quantities $\bar{\sigma}(t)$ as the volatility of the microscopic model and $\sigma(L,\bar{h}(t))$ as the according property assigned to the macroscopic eq.~(\ref{eq:macro}). In this paper all averages are performed over 30 time steps. The reasonable agreement allows for the conclusion that the system behaves mostly as an equilibrated system with the actual $\bar{h}$ and the use of eq.~(\ref{eq:macro}) is justified. 

Therefore, let us continue by investigating the properties of the microscopic model using the macroscopic model. Firstly we discuss the distribution of absolute returns. Since due to eq.~(\ref{eq:macro}) a single absolute return $|\Delta m|>0$ is distributed as
\begin{eqnarray*}
g(|\Delta m|) & = & \frac{2}{\sqrt{2\pi}\sigma\left(L,\bar{h}(t)\right)} \exp \left(\frac{-\Delta m^2}{2\sigma\left(L,\bar{h}(t)\right)^2}\right),
\end{eqnarray*}
the absolute return distribution of the whole time series is calculated as the average 
\begin{eqnarray}
\rho(|\Delta m|) & = & {\frac{\sqrt 2}{T\sqrt{\pi}}} \sum_{t=1}^T \frac{1}{\sigma\left(L,\bar{h}(t)\right)}\exp \left(\frac{-\Delta m^2}{2\sigma\left(L,\bar{h}(t)\right)^2}\right). \label{eq:returns}
\end{eqnarray}
In Fig.~\ref{fig:returns} the absolute return density is compared to the outcome of eq.~(\ref{eq:returns}) with reasonable agreement. So the power law distribution for the case $L=128,\ \alpha=20$ emerges due to the slowly but strongly varying $\sigma(t)$ as a superposition of Gaussian distributed returns with varying standard deviation for different times. For the larger system with $L=1024$ and $\alpha=60$ a double humped structure can be seen. The additionally plotted Gaussian with the typical volatility for the disordered phase emphasizes its origin in the increasing difference between returns in the ordered and the disordered phase with increasing system size. The differences at intermediate returns rely on equilibration effects around the phase transition, which are only present in the microscopic model. So the mechanism producing power laws is understood including its limitations for larger systems. 

Finally we use the macroscopic view to discuss the time scales of the system. Since the partition of the random walk in magnetization broadens with time proportional to $\Delta m \cdot t^{1/2}$, it eventually will arrive at the borders $\pm h_{\rm crit}/\alpha$. This defines a time scale $t\propto \left( \alpha \Delta m \right)^{-2} \propto L^3/\alpha^2$. This is in fact found in the distribution of return-to-zero-times, which describes the bull and bear market durations. The distribution shows a power law with an exponent about $1/2$, as can be found in the case of uncorrelated random walks as well. The cutoff of this power law is shown in Table~\ref{tab:times} for different system parameters with the finding $T_{\rm cutoff}\propto L^3/\alpha^2$, which is exactly the scaling we expected for the system's timescale with the macroscopic model. 
\begin{table}
\caption{\label{tab:times}Cutoff in the density of return to zero times.} 
\begin{ruledtabular} 
\begin{tabular}{lllll}
$L$ & 32 & 128 & 128 & 512\\
$\alpha$ & 10 & 10 & 20 & 80\\
$L^3/\alpha^2$ & 328 & 20972 & 5242 & 20972\\
$T_{\rm cutoff}$ & 400 & 30000 & 6000 & 20000\\
\end{tabular}
\end{ruledtabular} 
\end{table}

By characterizing the system state of the spin market model with the borderline length and by freezing the coupling to the global field, we found a phase transition between small and large magnetizations with different finite size scaling in the ordered and disordered phases. The scaling was understood with the border line length. We deduced a macroscopic equation for the price evolution and compared it to the microscopic model with reasonable agreement, which can be understood as follows: The system is equilibrated almost every time according to its smoothened mean field. So we proceeded by investigating the model dynamics with the macroscopic point of view. With this method we found the mechanism leading to power laws in the absolute returns, we found the reason for the appearance of a double humped structure in larger systems and we found and explained the system's time scale in terms of the bull and bear market durations. 
In summary, the mechanisms driving the spin market model are understood in a macroscopic way and the system parameters are characterized simultaneously.  
The model can be further adapted without losing the macroscopic description. So exceeding this paper, new model variants might be created in a controlled manner, e.g.\ by varying scaling properties or including multi-fractal elements \cite{calvet2001,lux2008} that real market mechanisms suggest.

\end{document}